
 \input harvmac

\def\np{Nucl. Phys. }
\def\pl{Phys. Lett. }
\def\hat{\widehat}
\def\pr{Phys. Rev. }
\def\prl{Phys. Rev. Lett. }

\def\ch{{\cal H}}

\def \cl{{\cal L}}

\Title{\vbox{\hbox{WIS/12/94/, EFI-94-67}
\hbox{\tt hep-th/9501024}}}
{{\vbox {\centerline{Universality in Two Dimensional Gauge Theory}
}}}

\centerline{\it D. Kutasov}
\smallskip\centerline
{Enrico Fermi Institute}
\centerline {and Department of Physics}
\centerline{University of Chicago}
\centerline{Chicago, IL60637, USA}
\vskip .2in
\centerline{\it and}
\vskip .2in
\centerline{\it A. Schwimmer}
\smallskip\centerline
{Department of Physics of Complex Systems}
\centerline{Weizmann Institute of Science}
\centerline{Rehovot, 76100, Israel}
\vskip .2in

\noindent
 We discuss two dimensional Yang -- Mills theories
 with massless fermions in arbitrary representations
 of a gauge group $G$. It is shown that the
 physics (spectrum and interactions) of the massive states
 in such models is independent of the detailed structure
 of the model, and only depends on the gauge group $G$
 and an integer $k$ measuring the total anomaly. The massless
 physics, which does depend on the details of the model,
 decouples (almost) completely from that of the massive one.
 As an example, we discuss the equivalence of QCD$_2$
 coupled to fermions in the adjoint, and fundamental
 representations.

\Date{12/94}
%

\newsec{Introduction}

It is well known
\nref\kz{V. G. Knizhnik  and A. B. Zamolodchikov, \np {\bf B247} (1984) 83.}%
\nref\gw{D. Gepner and E. Witten, \np {\bf B278} (1986) 493.}%
\nref\god{For a review see e.g.
P. Goddard and D. Olive, Int. Journal of Modern Physics {\bf1A}
(1986)303}%
\refs{\kz - \god}
that many conformal field theories (CFT) invariant under
a global symmetry based on a Lie algebra $G$
actually possess an infinite affine Lie algebra Kac Moody (KM)
symmetry $\hat G$, generated by conserved currents
$J^a(z)=\sum_n J^a_nz^{-n-1}$, satisfying
the operator product expansion / commutation relations:
\eqn\ope{\eqalign{
J^a(z)J^b(w)=&{k\delta^{ab}\over(z-w)^2}+i f^{abc}{J^c(w)\over z-w},\cr
[J^a_n,J^b_m]=&kn\delta_{n+m,0}\delta^{a,b}+i f^{abc}J^c_{n+m}.\cr}}
The integer $k$ is the level of the KM algebra, $f^{abc}$ the structure
constants of $G$. The Hilbert space of such CFT's can be described
as an in general infinite direct sum of highest weight representations
which are obtained by acting with current creation operators
$J^a_{n<0}$ on primaries of the KM algebra $\hat G$. CFT's can be
classified by $G$, $k$, and by the list of highest weight representations
appearing in a given model.

It is natural to ask what happens when such systems are coupled to non --
Abelian gauge fields, i.e. when one considers the quantum theory
based on the Lagrangian:
\eqn\egauge{\cl=\cl_{CFT}+\bar A J +A\bar J+{1\over g^2} F^2,}
where $F=\partial\bar A-\bar\partial A+[A,\bar A]$.
The main purpose of this note is to show that
the massive physics in such models is actually
{\it independent} of the particular realization
of $\hat G_k$ in the CFT. The massless sector described
by a coset model
\ref\gko{P. Goddard, A. Kent and D. Olive, \pl {\bf 152B}
(1985) 88.},
which carries the information
about the particular realization of $\hat G$, decouples almost
completely from the massive physics.

It is useful to illustrate some of the issues that arise in a solvable
example,                   the  generalized Schwinger model:
a set of $n$ right moving complex fermions $\psi_i$
with $U(1)$ charges $q_i$ ($i=1\cdots n$), and $m$ left
moving fermions $\bar\chi_j$
with charges $p_j$ ($j=1\cdots m$), coupled to a $U(1)$ gauge
field as in \egauge. This system was studied in
\ref\hal{I. Halliday, E. Rabinovici, A. Schwimmer and M. Chanowitz, \np {\bf
 B268} (1986)
413.}\
where it was shown that if the chiral anomaly vanishes, i.e.
\eqn\anom{k_l\equiv\sum_{i=1}^n q_i^2=\sum_{j=1}^m p_j^2\equiv k_r,}
the massive spectrum contains a single free scalar field with mass $M^2=g^2k$
($k=k_l=k_r$). The mass $M$ is independent of $q_i, p_j$ and $n,m$.
In addition there are some massless particles (see below).

To reveal the KM structure it is convenient to study the model
in a chiral gauge, $A^z\equiv A=0$. In  this gauge the action
\egauge\ takes the form:
\eqn\lca{\cl=\psi^\dagger_i\bar\partial\psi_i+\bar\chi_j^\dagger\partial
\bar\chi_j+\bar A J +{1\over g^2}(\partial\bar A)^2,}
where $J=\sum q_i\psi^\dagger_i\psi_i$. The ``Gauss Law'' constraint
$\delta\cl/\delta A=0$ enforces current conservation
$\partial\bar J+\bar\partial J=0$. Integrating out $\bar A$ leads
to
\eqn\lcb{\cl=\cl_{CFT}+g^2J{1\over \partial^2} J.}
Bosonizing $\psi_i,\;\bar\chi_j$ we find $n$ right moving
chiral scalars $H_i$ and $m$ left moving ones $\bar H_j$.
The interaction in \lcb\ gives mass to one combination
of the $H_i$, $\sqrt k H=\sum q_i H_i$, with $i\sqrt k\partial H=J$
($H$, $H_i$ are canonically normalized). Of course, a massive
particle may no longer be chiral. Current conservation
(and the related anomaly cancellation condition \anom) relates
one of the $m$ left moving scalars $\sqrt k \bar H=\sum p_j\bar H_j$
to $H$: $\bar H=H$. Thus the gauge theory \lca\ describes
one free massive and $n-1$ ($m-1$) free massless right (left)
moving particles. The flavour symmetry ($U(l)$ for charges appearing
$l$ times)
acts on the massless sector, which is decoupled from the
massive state $H(=\bar H)$. We will see below that in non abelian
gauge theories the massive and massless sectors
will in general be separately strongly interacting, but their
mutual decoupling is in fact general. As here, it is only the
massless sector (a coset CFT) that will carry the information
regarding the set of representations of $G$ one started with
(here $q_i, p_j, n,m$), and possible flavour symmetries.

To discuss the KM structure of massless gauge theories, it is further
convenient to employ Hamiltonian light -- front quantization of
\lca, \lcb\
\nref\LF{H. Pauli and S. Brodsky, \pr {D32} (1985) 1993, 2001;
K. Hornbostel, S. Brodsky and H. Pauli, \pr {D41} (1990) 3814.}%
\nref\LLF{R. Perry, A. Harindranath and K. Wilson, \prl {\bf 65}
(1990) 2959;
K. Wilson, T. Walhout, A. Harindranath, W. Zhang, R. Perry
and S. Glazek, \pr {\bf D49} (1994) 6720.}%
\refs{\LF, \LLF}.
One treats $x^-=z$ as a spatial coordinate, and $x^+=
\bar  z$ as ``time.'' Canonical quantization of \lcb\ then leads to a Hilbert
space spanned by creation operators of $\psi_i$ satisfying
the standard anticommutation relations\foot{It is easy to see that
the light -- cone gauge Gauss law mentioned above does not lead to constraints
on this Hilbert space.}, with
\eqn\ppm{\eqalign{
P^+=&\int dz\psi^\dagger\partial\psi,\cr
P^-=&\int dz({1\over\partial}J)^2.\cr}   }
The light -- cone momentum operator $P^+$
is diagonal on the $\psi$ Hilbert space;
one is looking for eigenstates of the light -- cone
Hamiltonian $P^-$. We see that it is natural to describe
the dynamics in terms of the KM structure,
since the Hamiltonian acts inside current blocks.
For a $U(1) $ gauge group, \lca, \lcb, current blocks are labeled
by the charge of the primary $J_0$. In the gauge theory
we have to impose the condition
\eqn\gl{J_0|{\rm phys}\rangle=0}
arising from fixing the residual $\bar z$ dependent
gauge symmetry $\delta\bar A=\bar\partial\epsilon(\bar z)$
of \lca. Therefore, physical states must all come from
the current block of the identity. In particular,
it is easy to verify using \ope, \ppm\ that the state
 $|p^+\rangle=J(-p^+)|0\rangle$
satisfies: $P^+|p^+\rangle=p^+|p^+\rangle$,
$P^-|p^+\rangle={g^2k\over p^+}|p^+\rangle$,
corresponding to the massive state mentioned
above as seen in the infinite momentum frame.

Note that the Hilbert space, which is the Fock
space of $\psi_i$, does not include the left moving
degrees of freedom $\bar \chi$. Mathematically, surfaces
with $x^+={\rm const}$ do not define a Cauchy problem
for $\bar\chi$. Physically, light -- front quantization
\ppm\ describes the Hilbert space seen by an observer
moving with the speed of light to the right. Such an observer
can see all massive particles, as well as all right moving massless
particles, but misses massless left moving ones.

It is important that while one misses the $m-1$
massless left moving scalars, one {\it does not}
miss any massive bound states of these massless
constituents. From the point of view of a stationary
frame, the point is that massive bound states
involve non -- trivial coupling between left and right
moving constituents, which in general in massless
$2d$ gauge theory occurs only through the anomaly, i.e.
only through the current sector\foot{Equivalently, in terms of
space -- time propagation, the only way right moving
massless particles can turn into
left moving ones is through pair annihilation into photons.}.
It is easy to show that the contact interaction
$\langle J(z)\bar J(0)\rangle=\pi\delta^2(z)$ in the free
CFT ensures current conservation in \lca.
Thus, while in light -- front quantization the physical
Hilbert space (the $\psi$'s) is only ``half''
of the full space of fixed time quantization (which also includes
the $\chi$'s), for {\it massive} physics the only aspect
of the $\bar\chi$'s that enters is the current $\bar J$,
which can be expressed in terms of $J$,
$\bar J=-{1\over\partial}\bar\partial J$. Hence, the $\psi$ Hilbert
space of light -- front quantization is sufficient to discuss
massive physics (as well as physics of the massless sector of
$\psi$). In fact light -- cone gauge together with quantization
in the infinite momentum frame makes (in two dimensions)
the decoupling between massive and massless, and left and right
moving particles most manifest.

Another important point that should be mentioned here concerns infrared
regularization. This is usually achieved by studying the
physical system on a spatial circle of finite radius.
The space -- time picture of bound state formation
mentioned above implies that such a regularization
is {\it inappropriate} for our purposes, since it
destroys the decoupling between left and right movers.
In addition to the coupling through the current sector,
a right moving particle can then interact with a left
moving one by going around the circle\foot{Of course,
as the radius of the circle goes to infinity
this coupling dissappears.}. Light front quantization
suggests a natural alternative: one can treat ``space'',
$x^-$ as compact, keeping ``time'', $x^+$ non -- compact.
This respects the decoupling between left and right movers, and we will use
it below.

The arguments above seem general, so in the next section
we will attempt to generalize them to the case of non -- abelian
gauge theories where they can be used to study the relation
between much less trivial theories.

\newsec{The decoupling theorem.}

Our previous comments suggest a strategy of dealing
with two dimensional Yang -- Mills theories with massless
quarks in arbitrary representations of the gauge group.
The purpose is to show that the physics of massive bound states
in such theories depends only on the gauge group $G$ and the KM
level $k$ (see \ope), and is independent of the detailed
representation content of the theory.

Consider the gauge theory  Lagrangian
\egauge\ with right handed quarks $\psi^{(r)}$
and left handed ones $\bar\chi^{(r^\prime)}$, in representations
$r$ and $r^\prime$ of $G$ respectively. Thus:
\eqn\ekin{\cl_{CFT}=\sum_r\psi^{\dagger(r)} \bar\partial\psi^{(r)}+
\sum_{r^\prime}\bar\chi^{\dagger(r^\prime)} \partial\bar\chi^{(r^\prime)}}
with the KM currents \ope\ given by:
\eqn\ecur{\eqalign{J^a=&\sum_r\psi^{\dagger(r)} \lambda^{a(r)}\psi^{(r)}\cr
\bar J^a=&\sum_{r^\prime}\bar\chi^{\dagger(r^\prime)}
\lambda^{a(r^\prime)}\bar\chi^{(r^\prime)},\cr}}
where $\lambda^{a(r)}$ are $G$ matrices in the representation
$r$, and summation over gauge indices has been suppressed.

The gauge theory \egauge\ is believed to be consistent
if the chiral anomaly vanishes, i.e. if the levels $k, \bar k$
of the right and left KM algebras \ecur\ coincide, $k=\bar k$.
Proceeding as in the Schwinger model, we choose the light -- cone gauge
$A^z=0$, and perform light -- front quantization with $z=$ space,
$\bar z=$ time. As discussed above, this makes the decoupling
of left and right movers most transparent. The Hilbert space
of the light -- front theory, which consists of $\psi$ creation operators
satisfying canonical anticommutation relations acting
on the vacuum, is just that of the right moving $\psi$ CFT
\ekin.

The Gauss law obtained by varying $\cl$ \egauge\ w.r.t.
$A^z$ enforces again
current conservation in the quantum theory,
\eqn\ct{\bar J=-{1\over\partial}\bar D J}
and contains no new information, simply stating the form
of $\bar J$ on the $\psi$ Hilbert space.
This complete decoupling of $\bar\chi$ is due (as before)
to the fact that in the frame moving to the right
with the speed of light, one can not see the left moving massless
particles.

The form of the light -- front Hamiltonian \ppm\ suggests
splitting the Hilbert space into $\hat G$ current blocks.
Putting $x^-=z$ on a circle, one finds:
\eqn\regul{P^-=\sum_{n=1}^\infty{1\over n^2}J^a_{-n}J^a_n.}
Thus, $P^-$ acts inside current blocks as before, and the problem of finding
the (massive) spectrum splits into decoupled CFT diagonalization problems
for the operator $P^-$ \regul\ on global $G$ singlets
(as in \gl) in the different current blocks.

What's most important for our purposes is that this description
of the light -- front dynamics is completely insensitive to the
properties of the left moving sector $\bar\chi$ \egauge, \ekin.
The only feature of $\bar\chi$ that has been used is anomaly
cancellation, $k=\bar k$. In particular, nothing prevents us
from replacing $\bar\chi$ by another massless CFT with the same
$\bar k$.
Gauge and Lorentz invariance of the chiral theory \egauge, \ekin\
allow us now to study the system in the two different light -- cone
gauges in appropriate infinite momentum frames and conclude
that $\psi$ too can be replaced by another set of fermions with the
same $k$ without changing the massive spectrum.
Since
the sets of representations $r$, $r^\prime$ are arbitrary, we arrive at
our main result: the spectrum depends only on the gauge group $G$
and the KM level (or total anomaly) $k$. It does not depend on the detailed
list of representations $r$ leading to that total anomaly $k$.
In particular, the spectra of the non chiral (left -- right symmetric)
theories corresponding to $\psi$, $\chi$ should coincide.

The essential ingredients of the above argument are decoupling
of left movers in a right moving infinite momentum frame and the
fact that
the spectrum of massive particles that can be boosted to
$v=+c$, which can be thought of as bound states of $\psi, J$
should be the same as that of particles that can be boosted to
$v=-c$ and can be thought of as bound states of $\bar\chi, \bar J$.

A few comments about this result are in order:

\noindent{} 1) In principle one should be able to identify different
states constructed  from $\psi$, $J$ in $A=0$ gauge with
$x^+(=\bar z)=$ time with different states in $\bar A=0$ gauge
with $x^-=$ time. However the argument described above does not
contain any detailed information about this mapping
in general. This is an interesting open
problem. We will study it in an example in section 3.

\noindent{} 2) ``Massive physics'' above includes the spectrum of
single particle states and their scattering amplitudes.

\noindent{} 3) As explained in the introduction, some massless
states are not seen in light -- front quantization
since the observer is moving with the speed of light. Massless
states can be studied  by familiar coset CFT techniques
which we discuss in section 4.

\noindent{} 4) For our results to be true, there must exist
a complete decoupling between massive physics, which is universal,
and massless physics as well as topological effects, which are model
dependent. This is indeed the case in two dimensional gauge theory
up to some global correlations to be discussed below.
As an example, flavour symmetry, which clearly depends on the representation
content can not be universal; consequently it must be carried
by the massless sector.

\noindent{} 5) Because of the above decoupling, physics of the massive states
may exhibit symmetries that are not apparent in the Lagrangian.
E.g. despite the fact that \egauge, \ekin\  is not in general
parity invariant, the parity violation is carried by the massless
sector.

\noindent{} 6) Due to the algebraic nature of the light -- front Hamiltonian
$P^-$
\regul\ it is clear that in all current blocks that are shared
by two different theories based on a given KM symmetry $\hat G_k$
there is an independent argument that physics will indeed be the same.
Thus, for example any two such gauge theories share the identity
current block, so at least part of the spectrum must be the same.
The issue, from this point of  view, is why the result is so general,
when different theories with $\hat G_k$ symmetry have in general distinct lists
 of
KM primaries. To understand how this may happen, we discuss in the next
section,
a non -- trivial example, $\hat{SU(N)}_N$ in the limit $N\to\infty$.

\newsec{An example: adjoint versus fundamental fermions in QCD$_2$.}

There has been some recent interest in the dynamics of two dimensional
gauge fields coupled to fermions in the adjoint representation of
$G=SU(N)$ in the large $N$ limit
\nref\DK{S. Dalley and I. Klebanov, \pr {\bf D47} (1993) 2517.}%
\nref\bdk{G. Bhanot, K. Demeterfi and I. Klebanov, \pr {\bf D48} (1993) 4980;
\np {\bf B418} (1994) 15.}%
\nref\kut{D. Kutasov, \np {\bf B414} (1994) 33.}%
\nref\bk{J. Boorstein and D. Kutasov, \np {\bf B421} (1994) 263.}%
\refs{\DK - \bk}.
The model exhibits a rich spectrum of bound states
with an infinite number of Regge trajectories
and may be a useful toy model for large $N$ QCD$_4$ as well as
QCD strings.
The level of the $\hat{SU(N)}$ KM symmetry \ope\ generated
by $J^{ab}=\psi^{ac}\psi^{cb}$, $a,b=1\cdots N$
is $N$. It is interesting to apply
our analysis to this system, since few exact results about it are available.
In particular, as a non -- trivial check  on our results one can
compare adjoint QCD$_2$ to a model of complex fermions $\psi^{\alpha a}$,
with a color index $a=1\cdots N_c=N$, and a flavour one $\alpha=1
\cdots N_f=N$ ($N$ flavours of fermions in the fundamental of
$SU(N_c)$). The $SU(N_c)$ current $J^{ab}=\psi^{\dagger a\alpha}
\psi^{\alpha b}$ generates a $\hat{SU(N)}_N$ KM algebra as well.
The arguments of the previous section would suggest that the two left --
right symmetric theories have the same massive spectrum for all $N$, and
in particular as $N\to\infty$ where the analysis simplifies.
We will now look at these spectra in some detail and attempt to compare
them.

\subsec{Adjoint fermions.}

It is convenient to put ``space'' $x^-=z$ on a circle
as described above, and take the fermions $\psi^{ab}$ to be
antiperiodic (Neveu -- Schwarz) around the circle. The Hilbert space
of global $SU(N)$ singlets \gl\ is then spanned by states of the form:
\eqn\hil{{1\over N^{l/2}}{\rm Tr} \left(\psi_{-r_1}\psi_{-r_2}
\cdots\psi_{-r_l}\right)|0\rangle;\;\;0<r_i\in Z+{1\over2};\;\;\;
l\geq2,}
where we take states with a single trace in \hil\ because
of the large $N$ limit. For massless constituents, the form of $P^-$
\regul\ suggests arranging the Hilbert space in a different way, according
to blocks of $\hat{SU(N)}$ KM. The diagonalization of $P^-$ splits
into decoupled problems for the different current blocks. To specify
the current blocks that appear, we need to determine the KM primaries
in the model. The two simplest current blocks are:

\noindent{} 1) The current block of the identity, with global $SU(N)$ singlets
of the form
\eqn\curid{{1\over N^l} {\rm Tr} \left(J_{-n_1}J_{-n_2}\cdots
J_{-n_l}\right)|0\rangle;\;\;0<n_i\in Z;\;l\geq 2.}

\noindent{} 2) The adjoint current block, with global $SU(N)$ singlets:
\eqn\curad{{1\over N^{l+{1\over2}}} {\rm Tr} \left(J_{-n_1}J_{-n_2}\cdots
J_{-n_l}\psi_{-{1\over2}}\right)|0\rangle;\;\;0<n_i\in Z;\;l\geq 1.}
More complicated highest weight states appear in products of $\psi$'s.
Explicitly, in the space of states of the form\foot{At finite $N$, it
is necessary to project out of these states
various contributions corresponding
to descendants.}:
\eqn\states{\prod_{i=1}^n\psi_{-{1\over2}}^{a_ib_i}|0\rangle}
one can find all representations of the form $T=\bar{S} R$ where the Young
tableaux of $S, R$ contain $n$ boxes, and $T$ is defined as follows:
if $R$, $S$ have columns of length $c_i$, $\tilde c_j$,
$i=1,\cdots, L$, $j=1,\cdots, \tilde L$, then the Young tableau
for $T$ has columns of length
\eqn\length{\cases{N-\tilde c_{\tilde L+1-i}& $i=1,\cdots ,\tilde L$\cr
                   c_{i-\tilde L}& $i=\tilde L+1,\cdots, \tilde L+L$\cr}}
Furthermore, due to antisymmetry of \states\ under interchange
of any two fermions it is clear that the Young tableau
for $S$ has to be the transpose of that for $R$: $S=R^t$.
Thus the highest weight representations of $\hat{SU(N)}$
that appear in this model are all representations of the form
$\bar{R}^tR$ with arbitrary $R$, with the length
of the first row $n_1$ and first column $c_1$ in the Young tableau
of $R$ satisfying $n_1+c_1\leq N$ (the unitarity
constraint of KM representation theory). A given $R$ corresponds
to a particular way of symmetrizing the indices $a_i$ in \states.
Each such highest weight state gives rise to a current block
which contains global $SU(N)$ singlets, since all the representations
involved are invariant under the center of $SU(N)$, $Z_N$; these can
be written analogously to \curid, \curad. One can diagonalize
$P^-$ \regul\ separately on the different current blocks.

In the large $N$ limit certain simplifications occur. All representations
with given $n$ in \states\ collapse to one, as far as single particle states
are concerned. The space of  potential  single particle states with
given $n$ is (schematically):
\eqn\curn{{\rm Tr} \left(J^{l_1}\psi J^{l_2}\psi\cdots J^{l_m}\psi
\right)|0\rangle.}
To illustrate this, consider the case $n=2$. The state
\eqn\ntwo{|\theta^{abcd}\rangle=\left(\psi^{ab}_{-{1\over2}}\psi^{cd}_{-{1\over2
 }}
-{1\over N}\delta^{bc} J^{ad}_{-1}+{1\over N}\delta^{ad}
 J^{cb}_{-1}+(a\leftrightarrow c)
\right)|0\rangle}
is a primary (compare to \states\ and the discussion following it), with $R$
 being the
two index symmetric representation. A simple global $SU(N)$ primary
is
\eqn\glsing{J^{da}_{-n_1}J^{bc}_{-n_2}|\theta^{abcd}\rangle.}
The terms in \glsing\ have the form ${\rm Tr}(J\psi J\psi)|0\rangle$,
$({\rm Tr} J\psi)({\rm Tr} J\psi)|0\rangle$, and ${1\over N} {\rm
 Tr}(JJJ)|0\rangle$.
Only the first of these can be in the single particle Hilbert space
as $N\to\infty$. The second term corresponds to two particles and the third is
 down by
$1/N$. The second representation with $n=2$ in \states\ corresponds
to $R=$ antisymmetric tensor; it differs from $\ntwo$ by some signs,
 corresponding
to antisymmetrization  rather than symmetrization  in $a,c$. At large $N$ these
 signs control
only the relative weight of the one and two trace terms in \ntwo. In general,
we
 see that
different representations $\bar{R}^tR$ where $R$ has $n$ boxes
differ by their projections on different multiparticle states. For given $R$
the
 states
are particular linear combinations of $1,2, \cdots, n$ trace states \curn. If
 one can
identify single trace states with single particle states,
the single particle spectrum should
split into sectors labeled by $n=0,1,2,\cdots$ \curn, in each of which
the Hamiltonian acts independently. We will soon see that actually
only the sectors with $n=0,1$ (i.e. \curid, \curad) give rise to single
particle states.

\subsec{$N$ flavours of fundamental fermions.}

Only representations invariant under the center of $SU(N)$
(which in this case are a subclass of all existing representations)
should be considered. In the baryon number zero sector, we have states
of the form (compare to \states):
\eqn\statesf{\prod_{i=1}^n
\psi^{\dagger a_i\alpha_i}_{-{1\over2}}
\psi^{\beta_i b_i}_{-{1\over2}}|0\rangle.}
It is easy to repeat the previous analysis
for this case. Since the flavour indices are at our
disposal, we can separately symmetrize $a_i$, $b_i$
in an arbitrary fashion. Hence, we find for given $n$
all highest weight states in representations of the form
$\bar{S}R$ with $R$, $S$ corresponding to arbitrary Young
tableaux with $n$ boxes. Of course, the (anti-) symmetry of
\statesf\ implies that given $R$, $S$ the transformation
properties under the flavour group are determined to
be $\bar{S}^tR^t$. Thus, we see that the list  of representations
of $SU(N_c)$ here is larger than the one obtained in the
adjoint case.

At large $N$ there is again a significant simplification.
Multiplying \statesf\ by products of currents to form global singlets
we see that all representations $\bar SR$ with $R$, $S$ containing $n$
boxes give linear combinations of states of the form:
\eqn\npart{\left[\psi^{\alpha_1a_1}\left(J^{l_1}\right)^{a_1b_1}
\psi^{\dagger b_1\beta_1}\right]
\left[\psi^{\alpha_2a_2}\left(J^{l_2}\right)^{a_2b_2}
\psi^{\dagger b_2\beta_2}\right]\cdots|0\rangle.}
We see that all such representations with given $n\geq1$ give rise to different
$n$ particle states. But since the diagonalization
of $P^-$ \regul\ is an algebraic problem insensitive to the
realization of the different current blocks, the conclusion
must be valid in any other representation of  $\hat{SU(N)}_N$ as well.
This implies that for $\hat{SU(N)}_{N\to\infty}$, the only
current blocks contributing single particle massive states are the identity
block \curid, and the adjoint block \curad.
Both appear in the two theories we are comparing, the identity
trivially \curid\ and the adjoint by replacing $\psi^{ab}\to\psi^{\dagger
a\alpha}\psi^{\beta b}$ in \curad.
The higher representations, which as we saw are different
in the two cases, contribute only multiparticle states, and so the difference
is unimportant.

\subsec{Comments.}

One interesting property of the mapping of states described above between
the adjoint and fundamental theories is that the fermionic states
in the adjoint current block in \curad\ are mapped into
bosonic ones in the $N$ flavour fundamental theory.
The reason for the discrepancy is clear: while a state
like ${\rm Tr} (J^n\psi)|0\rangle$
in the adjoint theory is a (fermionic) single particle state, in the
fundamental
theory it corresponds to $(J^n)^{ab}\psi^{\dagger b\alpha}\psi^{\beta a}
|0\rangle$ which has non trivial flavour content. Since $P^-$ \regul\
is completely insensitive to flavour, such states correspond
to two particle states where a massive particle interacting with the
color field is accompanied by a massless state sensitive only to
flavour. The two particle state is bosonic but it is difficult in general
to determine (independently) the statistics of the massive state
alone, since there is no
state in the Hilbert space of the theory with only the massive component
(see section 4 for more detail).
This pecularity is of no dynamical significance since, as mentioned
above, the massless state accompanying the massive one is a spectator
that does not participate in the dynamics.

The same can be said in general about the multiplicities
and flavour content of various states. Counting of states requires
taking into account the decoupled massless flavour degrees of freedom. After
doing that, the massive dynamics is found to be the same.

Note that we have found that many sectors that naively contain single particle
 states
in the large $N$ limit in adjoint QCD$_2$ (\curn\ with $n>1$) actually give
rise
to multi particle states. While it is possible to show this directly in the
 adjoint
model, it is more apparent in the fundamental representation \npart.

Finally, we haven't discussed here the non zero baryon number sectors\foot{
$U(1)_B$, like flavour is carried only by the massless
sector. Still these sectors give rise (in general) to new $SU(N_c)$ highest
weight representations.}. At large $N$, baryons become heavy (the ones
that are
exactly massless because they belong to the decoupled coset CFT
are of no interest to us), and therefore are outside
the scope of the present analysis. At finite $N$ there should presumably
be a mapping of the two models that includes them.
It is important to reiterate that while the analysis in this
section relied on certain simplifications of the large $N$ limit
of the gauge theories, the equivalence of the massive physics
should be a property of these theories for all values of $N$.

\newsec{The massless sector }
In this section we'll study in more detail the (de)coupling between the
massless
and massive sectors of $2d$ gauge theories.
The exact structure of the massless sector of the theory \egauge\ is most
 transparent
in the  $A_0$ gauge.
We use the Schrodinger representation, the left and right moving
fermions $\psi(x)$ and $\bar \chi (x)$ being operators on a circle
$ 0 \leq x \leq 2\pi $ with antiperiodic boundary conditions. The gauge
invariant Hilbert space is defined by projecting to the $0$ -- eigenvalue
states of the Gauss operator $I(x)$:
\eqn\gauss{I(x)=J_L(x)+J_R(x)-{[D_xE(x)]}^a,}
where
$J_L^a (x)  = \sum_r  \psi^{\dagger (r)} \lambda^{a(r)} \psi^{(r)}$ ,
$J_R^a (x) = \sum_{r'} \bar \chi^{\dagger (r')} \lambda^{a(r')}
\bar \chi^{r'}$ , $E(x)$ is the chromoelectric field , $D_x$
is the covariant derivative and $r$ and $r'$ run over the
set of left and right fermionic representations, respectively.

The Hilbert space of the fermions can be decomposed using the coset
construction  \gko:
\eqn\cosii{\ch_L=\sum_s \oplus  (\ch_s^c \otimes \ch_s^{\hat G ,k} )~~~
{}~\ch_R=\sum_{s'} \oplus  (\ch_{s'}^c \otimes \ch_{s'}^{\hat G ,k} )}
where $\ch_s^{\hat G ,k}$ is a highest weight $s$ representation
of the KM algebra $\hat G$ at level $k$, and $\ch_s^c$ are blocks
of the appropriate coset theory. The only property of the cosets we
will need in the following is that they accomodate an action
of the Virasoro algebra, i.e. they correspond to sets of massless
representations of the Lorentz group. Generally these
representations appear in nontrivial superpositions which do
not admit a simple interpretation in terms of massless fermions
or bosons.

{}From \gauss\ it is clear that the projection to $0$ -- eigenvalues
involves only the representations of $\hat G$, i.e. for fixed
$s$, $s'$ we should find the states in the product space
$\ch_s^{\hat G ,k} \otimes \ch_{s'}^{\hat G ,k}\otimes \ch^A =\ch^P$
which are annihilated by $I(x)$ \gauss, where $\ch^A$ is the Hilbert space
spanned
by the eigenstates of the gauge potential     $A^a(x)$. This defines
the gauge invariant Hilbert space $\ch^{GI}$. Since the gauge
potential is invariant under the center of $G$, solutions of
\gauss\
exist only if $s$ and $s'$ belong to the same element of the
center.

The hamiltonian in the $A_0=0$ gauge is :
\eqn\ham{\eqalign{H=&
                   {1\over k+h}
                  \int_0^{2\pi}\left[
      :  J_{L}^2(x)+J_{R}^2(x) : -2kA(x)
      ( J_{L}(x)-J_{R}(x))+
    2A(x)^2\right]  dx\cr
    +&{g^2\over 2}
      \int_0^{2\pi}
      (E(x))^2dx
+L_0^c +\bar L_0^c. \cr}
      }
$L_0^c, \bar L_0^c$  are the Virasoro generators acting
on the coset, and $h$ is the dual Coxeter number of $G$.
The hamiltonian decomposes into a part acting
on $\ch^{GI}$ and a decoupled piece acting on the coset. The states
in the left and right cosets remain therefore in the gauge
invariant Hilbert space and represent massless, decoupled
degrees of freedom.

Before studying their properties further
we have to make sure that there are no "accidentally massless"
states produced by the diagonalization of the $\ch^{GI}$ (coupled)
part. Since the coupled theory is superrenormalizable, the infrared
limit is obtained when the coupling flows to infinity. In this
limit the gauge field kinetic term disappears and the coupled part
of the theory becomes a ${\hat G}/{\hat G} $ topological
theory, i.e. there are no massless degrees of freedom left.

To complete the discussion of the physical Hilbert space
we need to consider the role of "big" gauge transformations
present when the left and right fermion representations
are invariant under elements $g_0$ of the center of $G$ \ref\los{
E. Witten, Nuovo Cimento {\bf 51A} (1979) 325, for a recent
discussion see F. Lenz, M. Shifman and M. Thies preprint
TPI-MINN-94/34-T, UMN-TH-1315-94, CTP-2391 (1994).}; e.g. for adjoint
QCD$_2$ the gauge group is $SU(N)/Z_N$; big gauge transformations
are labeled by elements of $Z_N$.

Consider a big gauge transformation $g(x)$ corresponding
to one of these elements :
\eqn\big{g(x+2\pi)=g(x) g_0}
Such a transformation leaves the states of the coset inert and
has a very simple action on $\ch^P$: besides transforming
$A^a (x) $
in the
usual way,
it induces an outer automorphism of the
KM algebra and therefore takes   a representation $s$ into
a representation $s^{g_0}$. The exact correspondence follows
from the permutation of the weights induced by $ g_0$ on the
extended Dynkin diagram of $\hat G$ \god.
The eigenstates of the big gauge transformations (i.e. "$\theta$-
vacua"  and states built on them  ) are linear combinations:
\eqn\toeta{\sum_{g_0} exp(if(\theta,g_0))~~
\ch_{s^{g_0}}^{\hat G ,k}
\otimes \ch_{s'^{g_0}}^{\hat G,k} \otimes \ch^{A^g} }
where the phases $f(\theta , g_0)$ form a representation
$\theta$ of the group composed by the $g_0$ elements.

One can ask whether the big gauge transformations \big\ lead to non -- trivial
dynamical effects (vacuum mixing). The answer is that they do not have any
 observable
consequences. The reason is that the hamiltonian \ham\
is diagonal in the representations of $\hat G$; it has no
off diagonal matrix elements between different terms in \toeta. Moreover
 diagonal elements
between representations related by $g_0$ will be the same. It
follows that physical quantities do not depend on $\theta$.
Therefore models having fermions with different behaviour under
the center can have the same massive physics. Note that
the $\theta$-independence arises without the presence of a global
symmetry in the lagrangian, unlike in four dimensions or in
the Schwinger model.

To illustrate the above procedure, consider the case discussed in Section $3$,
 i.e.
an $SU(N)$ colour group coupled to fermions in the adjoint
representation. In this case $g_0$ can be any of the elements
of the center
\eqn\cent{g_0=exp(i {2\pi\over  N}     n)~~~~~
n=0,1,2,\cdots,N-1. }
The automorphism corresponding to \cent\
with $n=1$,
takes the highest weight representation characterized by
the Young tableau $(n_1,n_2,...,n_{N-1})$ to the representation
corresponding to $(N-n_{N-1},n_1-n_{N-1},...,n_{N-2}-n_{N-1})$,
where $ n_i~~ i=1,...,N-1 $ denote the length of the rows. In
particular, by taking the $ R$ representations (in the notation
of Section 3) defined by Young tableaux with $n_1=k,~n_2=n_3=...=0 $
one generates all the representations which could appear.

In light -- front quantization the big gauge transformations are
realized in an amusing fashion:
on a given representation one should
simultaneously change the hamiltonian by the outer automorphism
acting on the currents and require singlet states under the global
charges corresponding to the transformed currents.

The general structure of the Hilbert space after diagonalizing
the hamiltonian , is:
\eqn\mass{\sum_{s,s'}\oplus  (\ch_s^c \otimes \ch_{s,s'}^{GI}
\otimes \ch_{s'}^c )}
where $\ch_s^c$ contain massless states and all the states
in $\ch_{s,s'}^{GI}$ are massive. From  \mass\
it is clear that
even though all the information about the massive states
is in $\ch_{s,s'}^{GI}$ , there is no factorization in the
mathematical sense between the massive and massless states: the
asymptotic state involves generally some massless component.
However since the massless states do not interact in
the Virasoro basis, the S-matrix is completely determined
by the massive part.

Generally the cosets $\ch_s^c$ represent
a conformal theory which cannot be simply described by free
fermions or bosons. If there is a continuous global symmetry
present, since it acts on Weyl fermions it is chirally conserved.
As a consequence it produces a KM algebra which will be part of
the chiral algebra of the coset. Therefore all the continuous flavour
symmetries are realized on the massless states. The massive parts
have charge zero under all the continuous symmetries,
including
baryon number. Discrete symmetries like fermion number
parity,
act on the massive states, however due to the structure
\mass\ they
have well defined values only when the massive states
are accompanied by the massless spectators.

The decoupled massless states are completely determined
by the coset construction  \cosii\
and they do not have any dynamics
unlike in four dimensions. It could happen that two different
theories would lead to identical cosets or to cosets having some
blocks in common, i.e. to the same theory in the infrared.
Moreover even when the cosets have an interpretation in terms of
massless bosons or fermions, the field operators of this objects
are usually nonpolynomial  in the basic fields , behaving in this
sense as "solitons". These facts have some superficial
resemblance to recently discovered features of four dimensional supersymmetric
QCD \ref\seib{N. Seiberg, preprint RU-94-82, IASSNS-HEP-94/98 (1994)}.

A simple example showing these features is provided by the
following two theories: color group $G=SU(N-2)$ with $N$
multiplets of complex fermions in the fundamental representation
and color group $G=SU(N+1)$ with $N-1$ multiplets in the fundamental
representation. The massless spectrum for both theories is given by
blocks described by ${N(N-1)}\over 2$ complex fermions. The baryon
number of the fermions is $2$ in units where the quarks have
baryon number $1$,
showing that they are solitons.

\newsec{Discussion.}

The main results obtained here are:

\noindent{} 1) The physics of massive bound states
in two dimensional gauge theory with massless quarks
is independent of most of the details of the representations
in which the quarks lie, and only depends on the gauge group
$G$ and the current algebra level $k$.

\noindent{} 2) There is complete decoupling between the
physics of the massive bound states, which is universal,
and that of the massless sector, which carries all the information
about flavour symmetries, baryon number, etc. In general,
this decoupling is somewhat obscured by the fact that states in the
Hilbert space have non -- trivial projections on both the
massless and the massive sectors, as is familiar from coset models.
To exhibit
the decoupling it is very useful to quantize the system on the light --
front. The whole structure is special to two dimensions and does not
seem to have a simple counterpart in $3+1$ dimensional QCD, where there
are strong interactions between massless and massive particles.

\noindent{} 3) The point of view presented here leads to
possible significant improvements in numerical studies of
QCD$_2$ with massless quarks. One learns that the dynamics splits
into current blocks, and furthermore in the example studied
in detail very few current blocks (two) lead to single particle
states. The resulting picture is very useful for thinking about boson --
fermion cancellation in adjoint QCD$_2$, which was studied in \kut, \bk\
on the basis of a similar cancellation that is generic in string
theory \ref\ks{D. Kutasov and N. Seiberg, \np {\bf B358} (1991) 600.}.
Here all bosons come from the identity sector \curid, while all fermions
come from the adjoint block \curad; it is clear that the spectrum of highly
excited states is going to be dominated by states with large $l$ in \curid,
\curad\
and hence will be the same for bosons and fermions. It therefore is plausible
that
${\rm Tr} (-)^F\exp(-\beta H)$ exhibits here the cancellations typical of
string
theory
\ks. We also saw that many sectors of the Hilbert space that naively
contribute single particle states, actually give rise to multi particle states
in the large $N$ limit. It is important to take this effect into account in
numerical
estimates of the density of highly excited states in adjoint QCD$_2$ \bdk.

We haven't succeeded in understanding the results regarding the above
universality in fixed time quantization, or in the path integral
formalism. Also, the chiral gauge theories \egauge\ play a central role.
It would be interesting to understand the results in different
ways. Clearly, it would be interesting to solve for the universal behaviour
of the massive sector of $2d$ gauge theories. The algebraic approach
followed here as well as the use of different realizations
of a given theory should prove useful for that.

\bigbreak\bigskip\centerline{{\bf Acknowledgements}}\nobreak

We thank T. Banks, J. Boorstein, M. Douglas, D. Gepner and
E. Martinec,
for discussions.
The work of D. K. was partially supported by a DOE OJI grant,
that of A. S. by BSF grant number 5360/2
and by the Minerva foundation.

\listrefs
\end